\documentclass[pdflatex,sn-mathphys]{sn-jnl}

\jyear{2023}%

\theoremstyle{thmstyleone}%
\theoremstyle{thmstyletwo}%

\theoremstyle{thmstylethree}%

\begin{document}

\title[Reduced seismic activity after mega-earthquakes]{Reduced seismic activity after mega earthquakes}


\author*[1,2,3]{\fnm{Yongwen} \sur{Zhang}}\email{zhangyongwen77@gmail.com}
\equalcont{These authors contributed equally to this work.}

\author[2]{\fnm{Maor} \sur{Elbaz}}\email{maor.elbaz@gmail.com}
\equalcont{These authors contributed equally to this work.}

\author[2]{\fnm{Shlomo} \sur{Havlin}}\email{havlins@gmail.com}

\author[3]{\fnm{Yosef} \sur{Ashkenazy}}\email{ashkena@bgu.ac.il}

\affil[1]{\orgdiv{Data Science Research Center, Faculty of Science}, \orgname{Kunming University of Science and Technology}, \orgaddress{\city{Kunming}, \country{China}}}

\affil[2]{\orgdiv{Physics Department}, \orgname{Bar-Ilan University}, \orgaddress{\city{Ramat-Gan}, \country{Israel}}}

\affil[3]{\orgdiv{Department of Environmental Physics}, \orgname{Ben-Gurion University of the Negev}, \orgaddress{\street{BIDR}, \city{Midreshet Ben-Gurion}, \postcode{8499000}, \country{Israel}}}


\abstract{
Mainshocks are often followed by increased earthquake activity (aftershocks).
According to the Omori-Utsu law, the rate of aftershocks decays as a power law over time. While aftershocks typically occur in the vicinity of the mainshock, previous studies have suggested that mainshocks can also trigger earthquakes in remote locations. Here we examine the earthquake rate in the days following mega-earthquakes (magnitude $\ge7.5$) and find that the rate is significantly lower beyond a certain distance from the epicenter compared to surrogate data. However, the remote earthquake rate after the strongest earthquakes (magnitude $\ge 8$) can also be significantly higher than that of the rate based on surrogate data. Comparing our findings to the global ETAS model, we find that the model does not capture the earthquake rate found in the data, hinting at a potential missing mechanism. We suggest that the diminished earthquake rate is due the release of global energy/tension subsequent to substantial mainshock events. This conjecture holds the potential to enhance our comprehension of the intricacies governing post-seismic activity.
}


%




\maketitle

\section{Introduction}\label{sec1}

Earthquakes pose a significant and perilous threat to humanity, given their highly destructive nature. While they are complex spatiotemporal phenomena, several empirical laws govern their behavior. Notable among them are the Gutenberg-Richter law \cite{Gutenberg1944a}, which describes the exponential decay of earthquake magnitude distribution, and the Omori-Utsu law \cite{Utsu1961,UTSU1972}, which elucidates the power-law decay of aftershock rates over time. Additionally, various scaling and power laws have been established concerning the distribution of waiting times between earthquakes \cite{Bak2002,Corral2003,Corral2003a,Zhang2019,Zhang2020}. Extensive research has been conducted on these laws, providing insights into earthquake activity models such as the Epidemic-Type Aftershock Sequence (ETAS) model \cite{Ogata1998}.

Researchers have been unable to identify definitive precursors that can be utilized to forecast the occurrence of large earthquakes in advance \cite{jordan2011operational}. However, the clustering of aftershocks suggests that earthquake timing is not entirely random \cite{UTSU1972,Ogata1988}. Moreover, a previous study demonstrated that consecutive interevent earthquake intervals exhibit correlated behavior rather than randomness. Specifically, shorter (longer) interevent intervals have a higher likelihood of being followed by shorter (longer) interevent intervals \cite{Livina2005}. Furthermore, the application of the Detrended Fluctuation Analysis (DFA) to interevent interval time series has revealed the presence of long-range (power-law) correlations and other memory measures within earthquake catalogs \cite{Lennartz2008,Fan2018b}. 
In their research, Zhang et al. \cite{Zhang2019,Zhang2020} studied the lagged interevent times and distances and found that there are significant correlations for short-time lags, but weaker correlations for longer time lags. 
From a different perspective, previous studies have indicated that the occurrence rate of seismic events (foreshocks) tends to increase prior to a mainshock \cite{ogata2017statistics, papazachos1975foreshocks, kagan1978statistical, jones1979some, console1993foreshock, Peng-Vidale-Ishii-et-al-2007:seismicity,gulia2019real,petrillo2021testing}, following the so-called ``inverse Omori law'' although this observation is not as robust as the Omori-Utsu law.

Another crucial aspect of earthquakes pertains to the mechanisms underlying their spatial propagation. Strong earthquakes often trigger a series of subsequent aftershocks in their vicinity due to increased static Coulomb stress \cite{reasenberg1992response,wyss2000change,Lippiello2009a}. However, the phenomenon of remote triggering, where seismic waves from a mainshock trigger earthquakes thousands of kilometers away, has been observed and explained by dynamic stress triggering \cite{tibi2003remote,Felzer,brodsky2014uses}. For instance, following the 8.6 Mw East Indian Ocean mainshock on April 11, 2012 (accompanied by a powerful 8.2 Mw aftershock), there was a significant global increase in the rate of earthquakes in remote areas \cite{pollitz201211}. Nevertheless, other studies have found no evidence supporting such remote triggering \cite{parsons2011absence,parsons2014global}. Moreover, a different study noted a decrease in earthquake activity in remote locations a few hours after a mainshock, attributing this decline to reduced detection capabilities caused by seismic waves \cite{iwata2008low}. From a physical standpoint, the global occurrence rate beyond the immediate aftershock zone could decrease following large earthquakes due to energy and stress release. This relaxation effect can be likened to the suppression of short waiting times observed in the Abelian sandpile model during a massive avalanche \cite{garber2009predicting}.

The objective of this study is to examine how the earthquake rate varies as a function of distance and time following mega-earthquakes. In order to achieve our objective, we have developed a statistical approach that enables us to measure the real occurrence of earthquakes after a mega-shock in comparison to a distribution of earthquake rates starting from randomly selected times. This is done while taking into account the same location, time window, and distance from the epicenter as the mega-earthquake. We statistically find that the earthquake rate proceeding mega-earthquakes and beyond a critical distance of about 100 km from the epicenter is significantly lower than the mean rate at the location of the mega-earthquakes. We propose that this critical distance can serve as a measure for determining the extent of aftershocks. We also find that the strongest mega-earthquakes are followed by reduced remote earthquake activity or by enhanced remote earthquake activity.

\section{Results}\label{sec2}

\begin{figure}[th!]%
\centering
\includegraphics[width=1\textwidth]{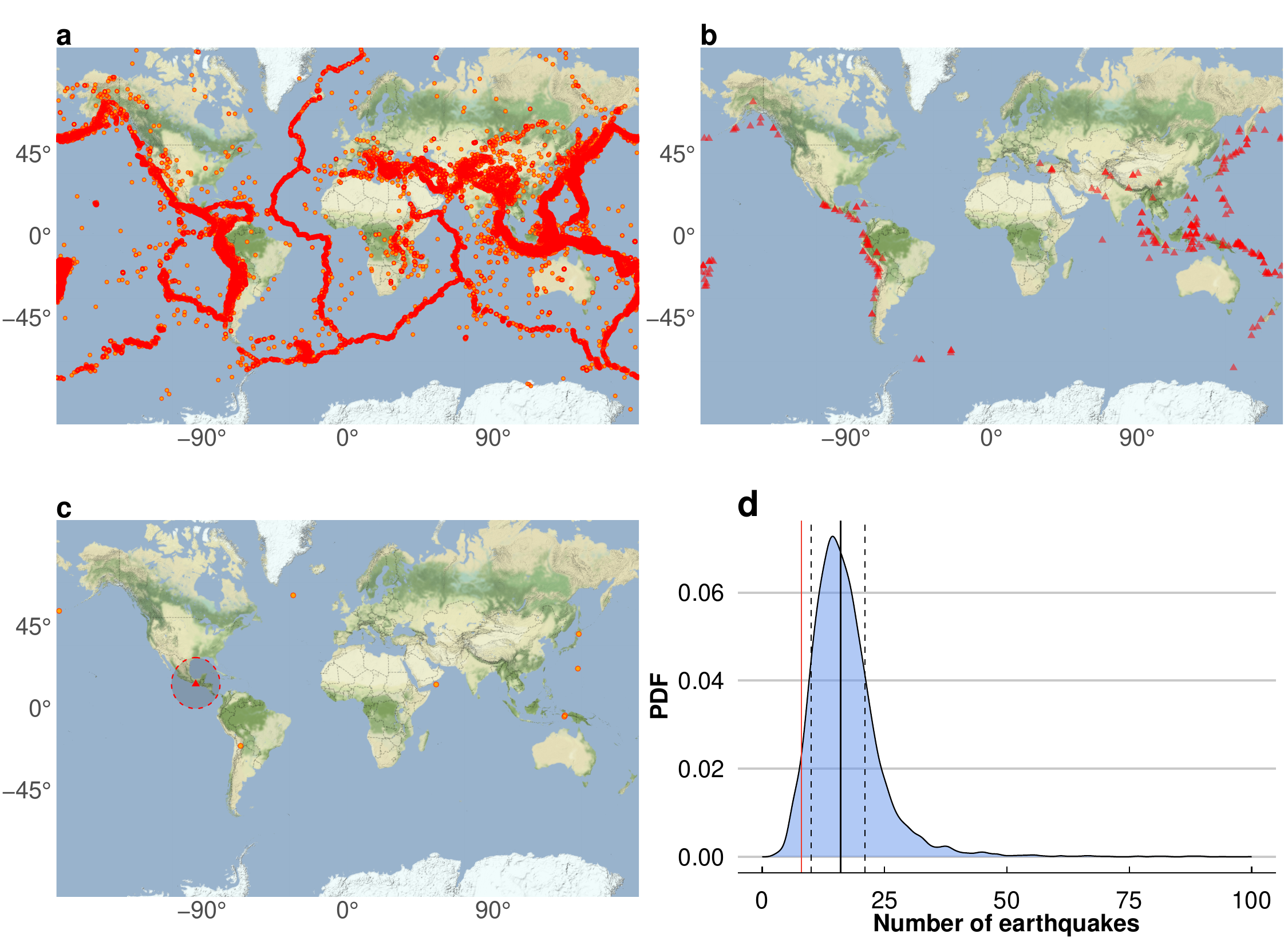}
\caption{\textbf{Demonstration of the proposed statistical method.}
\textbf{a} Spatial distribution of global earthquakes with a magnitude larger than or equal to 5.1 between May 1979 and May 2023. \textbf{b} Same as \textbf{a} but for only mega-earthquakes with a magnitude larger than or equal to 7.5. \textbf{c} An example of the 2017 Chiapas earthquake with a magnitude 8.2 (red triangle) and its following events (circles) above a certain distance (i.e., 500 km, marked by the shaded dashed circle) and within a time window of five days. \textbf{d} Probability Density Function (PDF) of the number of events at distances longer than 500 km from the location of the Chiapas earthquake and within 5 days for $10^4$ realizations of the surrogate data (i.e., $10^4$ realizations of randomly selected initial times). The dashed black vertical lines represent the 10$\%$ and 90$\%$ quantiles. The red line shows the observed number of earthquakes following the real 2017 Chiapas earthquake, which is below the 10\% quantile. The solid black vertical line represents the median.}\label{fig1}
\end{figure}

We utilized the comprehensive global earthquake catalog, between May 1979 and May 2023, setting a minimum magnitude threshold at 5.1 (for this value, the catalog can be regarded as complete; for details, see the Methods section).
The distribution of earthquake magnitudes 
closely conforms to the Gutenberg–Richter law, exhibiting a slope of -1.07$\pm$0.02 (Fig. S1). The spatial distribution of the events included in the catalog is depicted in Fig. \ref{fig1}a. Predominantly, seismic events occur along tectonic plate boundaries, propelled by the intricate interplay of Earth's crustal movement and compression.

We next consider earthquakes of exceptional magnitude, above or equal to 7.5, the common definition of mega-earthquakes \cite{kossobokov2022}. Fig. \ref{fig1}b depicts the geographical distribution of these mega-earthquakes, with prominent active regions including South America, Indonesia, and Japan. We demonstrate our analysis using the Chiapas earthquake, an 8.2 mega-earthquake, which occurred on September 7, 2017; Fig. \ref{fig1}c shows its location.
We analyzed the global earthquake occurrences (with magnitude $m \ge 5.1$) subsequent to this mega-earthquake within a five-day temporal window denoted as $T$, while excluding earthquakes within a radius ($R$) of $500$ km from the mega-earthquake's epicenter. We found a total of eight such earthquakes that followed the Chiapas mega-earthquake; see Fig. \ref{fig1}c.

We next study whether the earthquake rate after this mega-earthquake is larger, smaller, or equal to the mean earthquake rate associated with the mega-earthquake's location. We show the results in Fig. \ref{fig1}d, which depicts the Probability Density Function (PDF) of the number of earthquakes that occurred farther than 500 km from the epicenter of the 8.2 Chiapas mega-earthquake and within a time window of five days from randomly selected times. This PDF is utilized to create a null hypothesis concerning the rate observed after a mega-earthquake. If the observed rate falls within the PDF's confidence interval, typically between the 10\% and 90\% quantiles, then the null hypothesis is not rejected, and the rate is considered "normal." However, if the observed rate falls outside the confidence intervals, it is classified as either a low or high rate; see the Methods Section. Fig. \ref{fig1}d depicts the 10\% and 90\% quantiles (the dashed black vertical lines), as well as the observed rate (the solid red vertical line), indicating that the observed subsequent rate of the 2017 Chiapas mega-earthquake is low, below the 10\% quantile. This procedure helps to prevent possible geographical biases. 

\begin{figure}[th!]%
\centering
\includegraphics[width=0.9\textwidth]{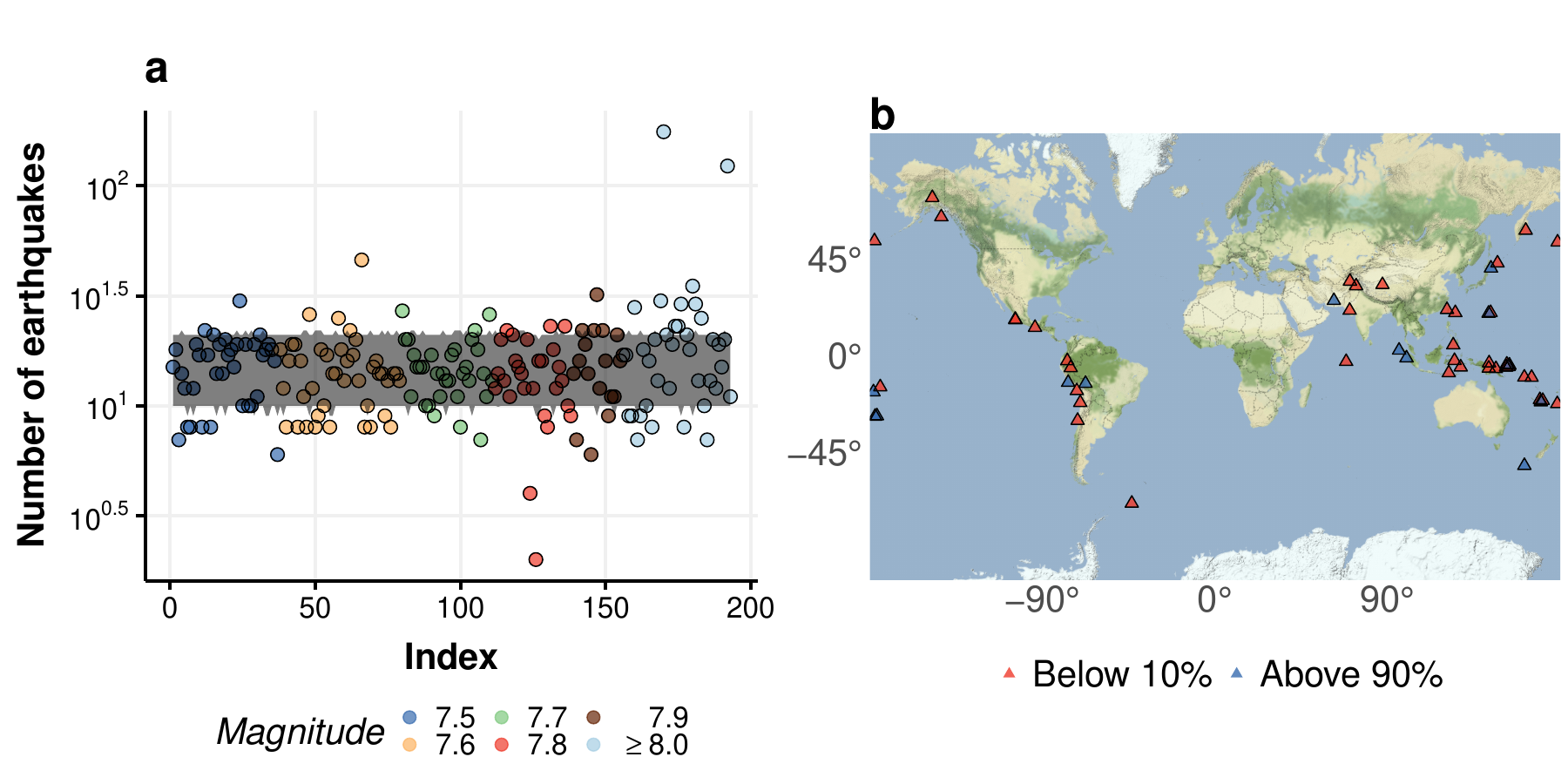}
\caption{\textbf{Significance of mega-earthquakes ($m\ge7.5$) and their spatial distribution.} \textbf{a} The number of earthquakes $n(T,R)$ in a time window of $T=5$ days and beyond the distance $R=500$ kilometers from the epicenter, following mega-earthquakes (colors represent their magnitudes). The grey shading indicates the 10$\%$ to 90$\%$ interval of the PDF of the number of events for the surrogate data (see Fig. \ref{fig1}d). \textbf{b} The spatial distribution of the mega-earthquakes shown in \textbf{a} that fall below the 10$\%$ quantile (red triangle) and above the 90$\%$ quantile (blue triangle).}\label{fig2}
\end{figure}

We next consider all the mega-earthquakes and study the earthquake rate that follows them. Fig. \ref{fig2}a depicts, for each mega-earthquake, the actual count of ensuing earthquakes spanning a five-day window ($T$) and exceeding a distance of 500 km ($R$) from the epicenter. Also plotted is the 10\%-90\% interval, based on the surrogate data (Fig. \ref{fig1}d). Notably, certain mega-earthquakes exhibit counts falling below the 10\% quantile, indicating reduced activity, while others surpass the 90$\%$ quantile, denoting increased activity (Fig. \ref{fig2}a). 

The spatial distribution of these statistically significant mega-earthquakes is displayed in Fig. \ref{fig2}b. It is apparent that the majority of these significant events are concentrated in the vicinity of active seismic zones. Additionally, mega-earthquakes followed by reduced activity outnumber those followed by increased activity. Quantitatively, the ratio, $r$, of mega-earthquakes followed by reduced activity is $r=0.194$ (uncertainty between 0.176 and 0.212), while those followed by increased activity hold a ratio of $r=0.078$ (see Methods). Within the context of the null hypothesis, we would anticipate a ratio close to 0.1, as predicted by our chosen 10\% and 90\% quantiles. The discernible departure from this expected ratio highlights the statistical significance of the ratio pertaining to reduced activity.

Following the above, we study the dependence of the magnitude of the mega-earthquake on the corresponding ratio, $r$. Our analysis (Fig. S2) reveals that no definitive correlation exists between the mega-earthquake's magnitude and the resultant ratio. For most magnitudes, the earthquake rate after mega-earthquakes is reduced, i.e., the ratio of mega earthquakes that fall below the 10\% quantile is significantly higher than 0.1, while the ratio of mega-earthquakes that fall above the 90\% quantile is significantly lower than 0.1. Yet, the strongest mega-earthquakes ($m\ge 8.0$) result in both ratios being significantly larger than the expected 0.1 ratio (Fig. S2), suggesting that such mega-earthquakes can either trigger long-distance worldwide earthquakes or lead to reduced worldwide seismic activity; see Table \ref{T1} for details regarding mega-earthquakes that were followed either by significantly reduced earthquake activity or by significantly increased earthquake activity. 

\begin{table}[!htb]
\caption{\label{T1}%
\textbf{The significant mega-earthquakes with magnitudes $m\ge 8.0$ for a time window of $T=5$ days and distance from the epicenter farther than $R=500$ km.} The last column shows the percentile of the number of events for the surrogate data smaller (smaller and equal) than the real number.}
\centering
\begin{tabular}{@{}llllll@{}}
\toprule
Region/location      & \begin{tabular}[c]{@{}l@{}}Date\\ (dd/mm/yyyy)\end{tabular} & Magnitude & Activity  & $n(T,R)$ & Percentile \\ \midrule
Kuril Islands    & 04/10/1994                                                  & 8.3       & Reduced*  & 10  & 9.2\% (14.0\%)   \\
Antofagasta      & 30/07/1995                                                  & 8.0       & Reduced   & 9   &  5.7\% (9.3\%)   \\
Colima-Jalisco   & 09/10/1995                                                  & 8.0       & Reduced   & 9     & 5.1\% (8.7\%)  \\
New Ireland      & 16/11/2000                                                  & 8.0       & Reduced   & 7     & 2.1\% (4.0\%)   \\ 
Solomon Islands  & 06/02/2013                                                  & 8.0       & Reduced   & 8    & 3.2\% (5.6\%)    \\ 
Illapel, Chile   & 16/09/2015                                                  & 8.3       & Reduced   & 7     & 1.5\% (3.0\%)   \\
Chiapas, Mexico          & 08/09/2017                                                  & 8.2       & Reduced   & 8     & 3.4\% (5.4\%)   \\
Peru             & 05/26/2019                                                  & 8.0       & Reduced   & 9   & 4.9\% (8.6\%)     \\ \midrule
Bolivia         & 09/06/1994                                                  & 8.2       & Increased & 29    & 93.6\% (94.6\%)    \\
Tokachi          & 25/09/2003                                                  & 8.2       & Increased & 29    & 94.6\% (95.2\%)    \\
Tasman Sea       & 23/12/2004                                                  & 8.1       & Increased & 176    & 100\% (100\%)   \\
Sumatra, Andaman & 26/12/2004                                                  & 9.1       & Increased & 123     & 100\% (100\%) \\
Kuril Islands    & 15/11/2006                                                  & 8.3       & Increased* & 25      & 89.2\% (90.8\%) \\
Peru             & 15/08/2007                                                  & 8.0       & Increased & 28      & 93.1\% (94.1\%)  \\
Fiji             & 19/08/2018                                                  & 8.2       & Increased & 35     & 97.4\% (97.7\%)   \\
Kermadec Islands & 04/03/2021                                                  & 8.1       & Increased & 30     & 94.9\% (95.6\%)  \\ \bottomrule
\end{tabular}
\footnotetext{`*' indicates marginal counts.}
\end{table}

\begin{figure}[th!]%
\centering
\includegraphics[width=0.9\textwidth]{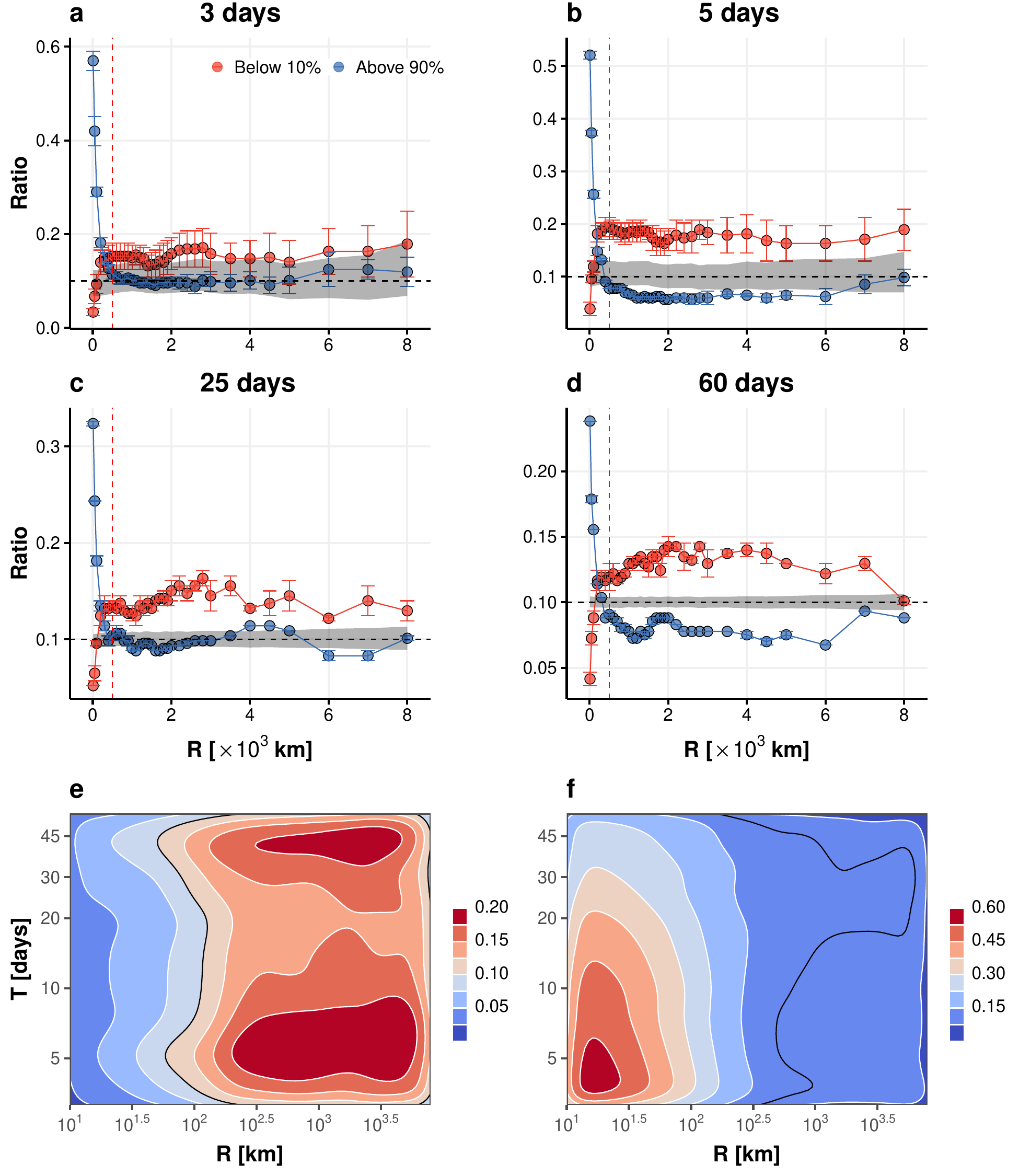}
\caption{
\textbf{The ratio of mega-earthquakes that fall below the 10\% quantile (red) and above the 90\% quantile (blue) as a function of the distance from the epicenter, $R$, and time window $T$.} \textbf{a} $T=3$ days, \textbf{b} $T=5$ days, \textbf{c} $T=25$ days, and \textbf{d} $T=60$ days. The dashed black horizontal line represents the ratio of the null hypothesis 0.1 and its uncertainty (grey shaded area). The dashed red vertical line indicates the transition distance of 500 km. The error bars are related to the uncertainty of the ratio as described in the Methods Section. The ratio of mega-earthquakes that \textbf{e} fall below the 10\% quantile and \textbf{f} above the 90\% quantile as a function of the distance from the epicenter, $R$, and time window $T$. The black contoured line represents the ratio $0.1$.}\label{fig3}
\end{figure}
 
Next, we explore the dependence of the ratio of mega-earthquakes that fall below (above) the 10\% (90\%) quantile as a function of both the distance from the epicenter $R$ (from 10 to 8000 km) and the time window $T$ (from 3 to 60 days). We first show these ratios as a function of the distance from the epicenter $R$ of the mega-earthquakes for different time windows $T$ (Fig. \ref{fig3}a-d). The effect of aftershocks is noticeable close to the mega-earthquakes' epicenters ($R\lesssim 250$) km, where the ratio is well off the expected 0.1 ratio, indicating significantly increased earthquake activity after the mega-earthquakes; i.e., the ratio of mega-earthquakes that fall below (above) the 10\% (90\%) quantile is much larger (smaller) than 0.1. 
For distances $R>500$ km, the ratios are more or less stable, indicating reduced earthquake activity after mega-earthquakes (ratio of $\sim0.15$ for ratios that fall below the 10\% quantile in comparison to the expected 0.1 ratio). Notably, the ratio of mega-earthquakes that fall below the 10\% quantile (red symbols in Fig. \ref{fig3}a-d) is more significant in comparison to the ratio of mega-earthquakes that fall above the 90\% quantile (blue symbols in Fig. \ref{fig3}a-d). 

The ratios of mega-earthquakes that fall below the 10\% quantile and above the 90\% quantile as a function of the distance, $R$, from the epicenter and time window $T$ are presented in Fig. \ref{fig3}e, f respectively. It is noticeable from Fig. \ref{fig3}e that the ratio of mega-earthquakes that fall below the 10\% quantile is not so sensitive to the time window $T$---this ratio is smaller than 0.1 for distances smaller than $\sim100$ km and higher than 0.1 otherwise, indicating enhanced earthquake activity after mega-earthquakes close to their epicenters (aftershocks) and reduced earthquake activity at farther distances.
Fig. \ref{fig3}f depicts the ratio of mega-earthquakes that fall above the 90\% quantile. Here the ratio is far above the expected 0.1 ratio for close distances ($R\lesssim 300$ km) and short time windows ($T\lesssim 30$ days), indicating aftershock activity which decays with distance and time \cite{Ogata1988,ogata2017statistics}. For large distances ($R>1000$ km), the ratio drops below the expected 0.1 ratio, indicating reduced earthquake activity after mega-earthquakes for distances greater than $\sim1000$ km. The transition across the 0.1 ratio is different for Fig. \ref{fig3}e ($\sim 100$ km) and Fig. \ref{fig3}f ($\sim 1000$ km), suggesting an aftershock extent for mega-earthquakes of about 500 km.

To substantiate the robustness of our findings, we not only evaluate the seismic activity significance based on the 10$\%$ and 90$\%$ quantiles, but also explore alternative metrics such as the median, average, and multiples of the average, as detailed in the Supplementary Figures S3-S6. Notably, all analyses consistently indicate a reduced earthquake activity after the occurrence of mega-earthquakes. Moreover, the results for different magnitude thresholds, 5.0 and 5.2, also show similar behavior in Figs. S7 and S8.      

\begin{figure}[th!]%
\centering
\includegraphics[width=0.9\textwidth]{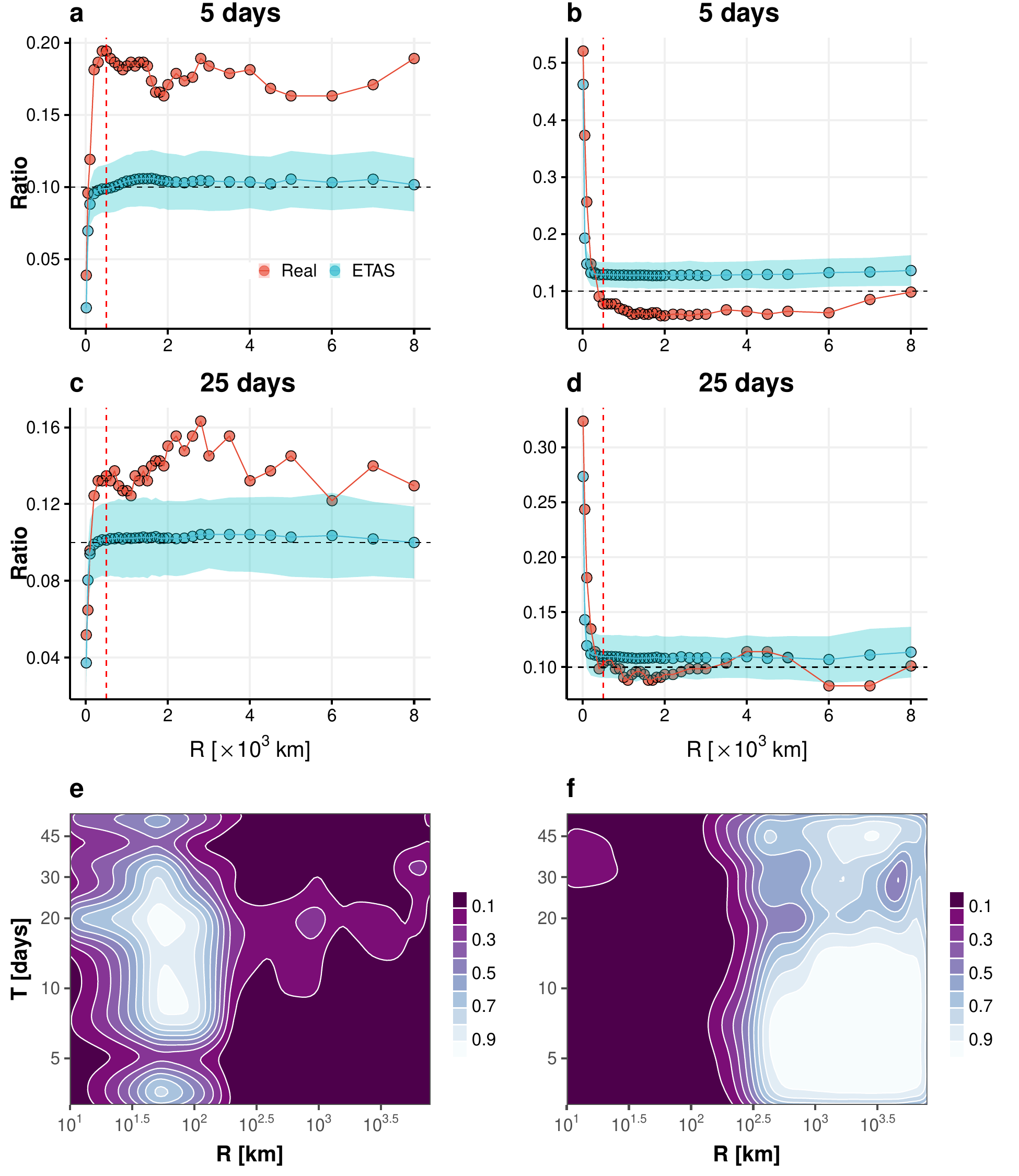}
\caption{\textbf{Comparison of the ratio based on the real global earthquake catalog and the synthetic ETAS model catalogs.} The ratio of mega-earthquakes that fall below the 10\% quantile for \textbf{a} $T=5$ days and \textbf{c} $T=25$ days as a function of distance from the epicenter $R$, for the real catalog (red symbols) and for the ETAS model synthetic catalog (cyan symbols). The ratio of mega-earthquakes that fall above the 90\% quantile for \textbf{b} $T=5$ days and \textbf{d} $T=25$ days as a function of distance from the epicenter $R$. The dashed black horizontal line represents the ratio of the null hypothesis 0.1. The dashed red vertical line indicates the transition distance of 500 km. The ratio of the ETAS model is averaged over 50 independent realizations, and the shading indicates the standard deviation.
The fraction of ETAS model realizations with a ratio exceeding that of the real data for the ratio of \textbf{e} mega-earthquakes that fall below the 10\% quantile and \textbf{f} above the 90\% quantile as a function of the distance from the epicenter, $R$, and time window $T$. A value of 0.5 indicates the similarity of the ETAS model to the real catalog---it is apparent that the ratios based on the ETAS model catalog are different than those of the real catalog, for the vast majority of $R$ and $T$.}\label{fig4}
\end{figure}

The global ETAS model \cite{Ogata1988, Ogata1998, Zhuang2010, Zhuang2012} was proposed to model the statistical properties (the spatiotemporal clustering) of the global earthquake catalog; see the Methods Section. We next compare the results described above, which are based on the real global catalog, to the synthetic catalog of the ETAS model.
The ETAS model reproduces the generation of a sequence of aftershocks following a major seismic event, which serves as the foundation for interlinking earthquakes in the model. Notably, from a mechanistic standpoint, the ETAS model lacks any intrinsic processes that model instances of reduced seismic activity. In Fig. \ref{fig4}a and c (Fig. \ref{fig4}b and d), we juxtapose the ratio of mega-earthquakes that fall below (above) the 10\% (90\%) quantile, for the real global earthquake catalog and for the synthetic global ETAS model catalogs.
As seen, the global ETAS model does not reproduce the reduced earthquake rate after mega-earthquakes for distances beyond several hundreds of km; see Fig. \ref{fig4}. This disparity arises due to the ETAS model's inability to replicate long-distance suppression of earthquake activity. However, for distances smaller than 500 km, the results based on the ETAS model synthetic catalogs align more closely with the results based on the real global catalog, attributed to the ETAS model's proficiency in representing aftershock clustering dynamics.
The ETAS model's ratio of mega-earthquakes that fall above the 90\% quantile slightly surpasses the 0.1 ratio for distances greater than several hundreds of km, particularly for $T=5$ days. This discrepancy arises due to the possibility of long-distance triggering in the ETAS model as outlined in Eq. \ref{eq4} in the Methods Section; see \cite{Ogata1988, Ogata1998, Zhuang2010, Zhuang2012}. 

We quantify the fraction of ETAS model realizations with a ratio exceeding that of the real data in a spatiotemporal context, as depicted in Fig. \ref{fig4}e and f. This ratio should be around 0.5 when the ETAS model results are comparable to the results of the real earthquake catalog. At short distances, nearly 90$\%$ of the ETAS realizations exhibit a ratio of reduced activity after mega-earthquakes, surpassing that of the real data (Fig. \ref{fig4}e). This discrepancy signifies that the ETAS model underestimates the rate of aftershocks in the proximity of the mega-earthquakes' epicenter. However, this underestimation does not significantly impact the results at greater distances. For extended distances, the fraction below 10\% in Fig. \ref{fig4}e underscores the substantial difference in the ratio of reduced activity between the real data and the ETAS model. In contrast, the ratio of mega-earthquakes that fall above the 90\% quantile for real data is considerably lower than the corresponding ratio from the ETAS model, as shown in Fig. \ref{fig4}f. Following the above, we can conclude that the ETAS model does not reproduce the results of decreased earthquake activity after mega-earthquakes. 

\section{Discussion and Conclusion}\label{sec3}

In this present study, we investigated the earthquake rate after the occurrence of mega-earthquakes (with a magnitude larger than or equal to 7.5). 
We developed several statistical methods to investigate this question, all resulting in reduced earthquake activity beyond a distance of several hundred km from the mega-earthquakes' epicenter. For smaller distances, increased earthquake activity is observed after the occurrence of mega-earthquakes and is attributed to aftershocks. The transition distance of about 500 km is suggested to be a length scale of aftershocks following mega-earthquakes with a magnitude larger than or equal to 7.5. We compare the results based on the real global catalog to the results obtained based on the global ETAS model catalogs and find that the ETAS fails to reproduce the results based on the real global earthquake catalog. This suggests that some key processes are missing in the ETAS model. For the strongest mega-earthquakes (with magnitude $m\ge 8$), we find that some were followed by significantly reduced remote earthquake activity while others resulted in significantly increased remote earthquake activity; the latter can be associated with remote triggering as was reported in previous studies. This observation should be verified more thoroughly as the total number of such mega-earthquakes is not large (37 events). 

The stress of faults could be transferred to more remote areas along tectonic plate boundaries or fault systems after a mega-earthquake occurs \cite{Scholz2019}. We conjecture that the stress transfer can change the stress state of remote faults, either promoting or inhibiting seismic activity in remote areas. These areas may experience a period of seismic quiescence, characterized by reduced seismic activity, following a mega-earthquake. This quiescent period can be a natural response to the redistribution of stress. Our finding provides researchers with an opportunity to study the geological processes between remote areas responsible for stress accumulation and release in greater detail. This can enhance our understanding of seismic hazards and improve earthquake prediction models.


\section{Methods}\label{sec4}

\subsection{Data}

We analyzed seismic events with a magnitude of 5.1 or higher (mW, mb, and ms) from May 1979 to May 2023 (44 years), using the United States Geological Survey (USGS) catalog (\url{https://www.usgs.gov/}). There were 57,085 earthquakes, with 193 events having a magnitude of 7.5 or higher, and 37 events with a magnitude of 8 or higher. The earthquakes included in the catalog followed the Gutenberg-Richter law (Fig. S1).

\subsection{Significance test}
It is widely recognized that mega-earthquakes have the potential to initiate a cascade of aftershocks within a condensed temporal window and spatial region. Here we examined post-mega-earthquake seismic activity, focusing specifically on regions located beyond a defined distance from the epicenter, and within a designated time interval. The metric utilized is the count of earthquakes, denoted as $n(T,R)$, occurring within a time window of $T$ days and at a distance exceeding $R$ km from the epicenter. Notably, earthquakes occurring within the spatial confines of the immediate aftershock region are disregarded.  

The null hypothesis posits that the count $n(T,R)$ and the timing of the mega-earthquake are statistically independent. Consequently, surrogate data is generated under the framework of this null hypothesis. For each mega-earthquake, a random initial time is selected from within the time span of the global catalog, while the geographical coordinates remain as for the original mega-earthquake; the latter constraint is aimed at avoiding possible spatial biases. The random selection of times is repeated many ($10^4$) times, yielding the surrogate earthquake count denoted as $n^\prime(T,R)$. We then construct the PDF characterizing the distribution of $n^\prime(T, R)$. Using this PDF, we establish the significance levels at 10\% for the lower bound and 90\% for the upper bound. Should the actual count $n(T,R)$ fall below (above) the 10\% (90\%) quantile, the mega-earthquake is considered to be a significant event, signifying that it was followed either by reduced or increased seismic activity, respectively.

\subsection{The ratio of significant events}
In the context of a sequence of mega-earthquakes, we determine the ratio $r=N_s/N_m$, where $N_s$ corresponds to the count of significant mega-earthquakes (i.e., mega-earthquakes that fall below/above the 10\%/90\% quantile), and $N_m$ denotes the overall count of mega-earthquakes. Notably, the quantile values of 10$\%$ and 90$\%$ may not be integral, necessitating the utilization of either the floor or ceiling integer values of these quantiles for comparison with the actual counts. Consequently, the count $N_s$ may exhibit variability when these floor or ceiling integers are applied, thus introducing uncertainty into the derived ratio. This uncertainty is especially large for short time windows that result in a low number of counts after the mega-earthquakes. To mitigate this uncertainty, we adopt the average of the counts obtained using both the floor and ceiling integers for the calculation of the ratio. The associated error bar is established as the difference between these two counts.

\subsection{The global ETAS model}

The ETAS model, a space-time stochastic point process, is employed to simulate synthetic earthquake catalogs \cite{Ogata1988, Ogata1998}. This model simulates seismic activities based on a rate function $\lambda$, defined at a location $(x, y)$ and time $t$, conditioned on the prior history $H_t$:
\begin{equation}\label{eq1}
\lambda(x, y, t \mid H_t) = \mu(x, y) + \sum_{t_i < t}{k(M_i)g(t - t_i)f(x - x_i, y - y_i, M_i)}.
\end{equation}
Here, $t_i$ indicates the time of past events, and $M_{i}$ represents their magnitude. The magnitude of each event ($\ge M_0$ where $M_0=5.1$ is the magnitude threshold of the catalog) is generated independently following the Gutenberg--Richter distribution with a magnitude-frequency parameter of $b=1$. The background intensity $\mu(x, y) = \mu_{0}u(x, y)$ at location $(x, y)$ is governed by the spatial PDF of background events denoted by $u$, estimated using the approach outlined in Refs. \cite{Zhuang2010, Zhuang2012}. $\mu_{0}$ stands for the background rate of seismic events on a global scale.

The magnitude-dependent triggering capability is formulated as
\begin{equation}\label{eq2}
k(M_i) = A\exp(\alpha(M_i - M_0)),
\end{equation}
where $A$ represents the rate of earthquakes at zero time lag, and $\alpha$ denotes the productivity parameter. The temporal decay of triggered events is described by the Omori law,
\begin{equation}\label{eq3}
g(t - t_i) = \left(1 + \frac{t - t_i}{c}\right)^{-p},
\end{equation}
where $c$ and $p$ are Omori law parameters. Spatial clustering of aftershocks is introduced through the spatial kernel function $f(x - x_i, y - y_i, M_{i})$ \cite{Zhuang2012},
\begin{equation}\label{eq4}
f(x - x_i, y - y_i, M_{i}) = \frac{q - 1}{\pi \zeta^2}\left(1 + \frac{(x - x_i)^{2} + (y - y_i)^{2}}{\zeta^2}\right)^{-q}.
\end{equation}
Here, $\zeta = D\exp[\gamma_m(M_{i} - M_{0})]$ signifies that the distances between triggering and triggered events depend on the magnitudes of the triggering events. Parameters $q$, $D$, and $\gamma_m$ are estimated parameters. We note that the ETAS model exhibits minimal long-range spatial correlation ($\gg$ $D$). The parameter values are estimated through the expectation-maximization algorithm, as detailed in Refs. \cite{veen2008estimation, nandan2021seismicity}, and summarized in Table \ref{T2}.

\begin{table}[!htb]
\caption{\label{T2}%
Estimated parameters of the global ETAS model.}
\centering
\begin{tabular}{ccccccccc}
\toprule
\textrm{}&
\textrm{$\mu_0$ (day$^{-1}$)}&
\textrm{$A$}&
\textrm{$\alpha$}&
\textrm{$p$}&
\textrm{$c$ (days)}&
\textrm{$q$}&
\textrm{$D$ (km)}&
\textrm{$\gamma_m$}\\
\midrule
\textrm{Parameters} & $2.8$ & $1.38$ & $1.70$ & $1.06$ & $0.0083$ & $2.03$ & $9.2$ & $0.50$ \\
\bottomrule
\end{tabular}
\end{table}

\backmatter

\bmhead{Acknowledgments}
We are grateful for the financial support provided by the Israel Ministry of Energy, the EU H2020 project RISE, the Israel Science Foundation (grant no. 189/19), DTRA, the Pazy Foundation, the joint China-Israel Science Foundation (grant no. 3132/19), the National Natural Science Foundation of China (grant no. 12305044) and the BIU Center for Research in Applied Cryptography and Cyber Security. We thank Golan Bel for the helpful discussions.

\section*{Declarations}

\begin{itemize}
\item Competing Interests: The authors declare that they have no
competing financial interests.
\item Data availability: The global earthquake catalog can be downloaded from the United States Geological Survey (USGS) (\url{https://www.usgs.gov/}). 
\item Code availability: The code is available on request from the authors.     
\end{itemize}



\begin{thebibliography}{10}
\expandafter\ifx\csname url\endcsname\relax
  \def\url#1{\texttt{#1}}\fi
\expandafter\ifx\csname urlprefix\endcsname\relax\def\urlprefix{URL }\fi
\providecommand{\bibinfo}[2]{#2}
\providecommand{\eprint}[2][]{\url{#2}}

\bibitem{Gutenberg1944a}
\bibinfo{author}{Gutenberg, B.} \& \bibinfo{author}{Richter, C.~F.}
\newblock \bibinfo{title}{{Frequency of Earthquakes in California}}.
\newblock \emph{\bibinfo{journal}{Bull. Seismol. Soc. Am.}}
  \textbf{\bibinfo{volume}{34}}, \bibinfo{pages}{185--188}
  (\bibinfo{year}{1944}).

\bibitem{Utsu1961}
\bibinfo{author}{Utsu, T.}
\newblock \bibinfo{title}{{A statistical study on the occurrence of af-
  tershocks}}.
\newblock \emph{\bibinfo{journal}{Geophys. Mag.}}
  \textbf{\bibinfo{volume}{30}}, \bibinfo{pages}{521--605}
  (\bibinfo{year}{1961}).

\bibitem{UTSU1972}
\bibinfo{author}{Utsu, T.}
\newblock \bibinfo{title}{{Aftershocks and Earthquake Statistics (3) : Analyses
  of the Distribution of Earthquakes in Magnitude, Time and Space with Special
  Consideration to Clustering Characteristics of Earthquake Occurrence(1)}}.
\newblock \emph{\bibinfo{journal}{J. Fac. Sci. Hokkaido Univ. Ser. 7,
  Geophys.}} \textbf{\bibinfo{volume}{4}}, \bibinfo{pages}{1--42}
  (\bibinfo{year}{1972}).

\bibitem{Bak2002}
\bibinfo{author}{Bak, P.}, \bibinfo{author}{Christensen, K.},
  \bibinfo{author}{Danon, L.} \& \bibinfo{author}{Scanlon, T.}
\newblock \bibinfo{title}{{Unified Scaling Law for Earthquakes}}.
\newblock \emph{\bibinfo{journal}{Phys. Rev. Lett.}}
  \textbf{\bibinfo{volume}{88}}, \bibinfo{pages}{178501}
  (\bibinfo{year}{2002}).

\bibitem{Corral2003}
\bibinfo{author}{Corral, {\'{A}}.}
\newblock \bibinfo{title}{{Local distributions and rate fluctuations in a
  unified scaling law for earthquakes}}.
\newblock \emph{\bibinfo{journal}{Phys. Rev. E}} \textbf{\bibinfo{volume}{68}},
  \bibinfo{pages}{035102} (\bibinfo{year}{2003}).

\bibitem{Corral2003a}
\bibinfo{author}{Corral, {\'{A}}.}
\newblock \bibinfo{title}{{Long-Term Clustering, Scaling, and Universality in
  the Temporal Occurrence of Earthquakes}}.
\newblock \emph{\bibinfo{journal}{Phys. Rev. Lett.}}
  \textbf{\bibinfo{volume}{92}}, \bibinfo{pages}{108501}
  (\bibinfo{year}{2004}).

\bibitem{Zhang2019}
\bibinfo{author}{Zhang, Y.} \emph{et~al.}
\newblock \bibinfo{title}{{Scaling laws in earthquake memory for interevent
  times and distances}}.
\newblock \emph{\bibinfo{journal}{Phys. Rev. Res.}}
  \textbf{\bibinfo{volume}{2}}, \bibinfo{pages}{013264} (\bibinfo{year}{2020}).

\bibitem{Zhang2020}
\bibinfo{author}{Zhang, Y.} \emph{et~al.}
\newblock \bibinfo{title}{{Improved earthquake aftershocks forecasting model
  based on long-term memory}}.
\newblock \emph{\bibinfo{journal}{New J. Phys}} \textbf{\bibinfo{volume}{23}},
  \bibinfo{pages}{042001} (\bibinfo{year}{2021}).

\bibitem{Ogata1998}
\bibinfo{author}{Ogata, Y.}
\newblock \bibinfo{title}{{Space-Time Point-Process Models for Earthquake
  Occurrences}}.
\newblock \emph{\bibinfo{journal}{Ann. Inst. Stat. Math.}}
  \textbf{\bibinfo{volume}{50}}, \bibinfo{pages}{379--402}
  (\bibinfo{year}{1998}).

\bibitem{jordan2011operational}
\bibinfo{author}{Jordan, T.~H.} \emph{et~al.}
\newblock \bibinfo{title}{Operational earthquake forecasting. state of
  knowledge and guidelines for utilization}.
\newblock \emph{\bibinfo{journal}{Ann. Geophys.}}
  \textbf{\bibinfo{volume}{54}}, \bibinfo{pages}{361--391}
  (\bibinfo{year}{2011}).

\bibitem{Ogata1988}
\bibinfo{author}{Ogata, Y.}
\newblock \bibinfo{title}{{Statistical Models for Earthquake Occurrences and
  Residual Analysis for Point Processes}}.
\newblock \emph{\bibinfo{journal}{J. Am. Stat. Assoc.}}
  \textbf{\bibinfo{volume}{83}}, \bibinfo{pages}{9--27} (\bibinfo{year}{1988}).

\bibitem{Livina2005}
\bibinfo{author}{Livina, V.~N.}, \bibinfo{author}{Havlin, S.} \&
  \bibinfo{author}{Bunde, A.}
\newblock \bibinfo{title}{{Memory in the Occurrence of Earthquakes}}.
\newblock \emph{\bibinfo{journal}{Phys. Rev. Lett.}}
  \textbf{\bibinfo{volume}{95}}, \bibinfo{pages}{208501}
  (\bibinfo{year}{2005}).

\bibitem{Lennartz2008}
\bibinfo{author}{Lennartz, S.}, \bibinfo{author}{Livina, V.~N.},
  \bibinfo{author}{Bunde, A.} \& \bibinfo{author}{Havlin, S.}
\newblock \bibinfo{title}{{Long-term memory in earthquakes and the distribution
  of interoccurrence times}}.
\newblock \emph{\bibinfo{journal}{EPL}} \textbf{\bibinfo{volume}{81}},
  \bibinfo{pages}{3--7} (\bibinfo{year}{2008}).

\bibitem{Fan2018b}
\bibinfo{author}{Fan, J.} \emph{et~al.}
\newblock \bibinfo{title}{{Possible origin of memory in earthquakes: Real
  catalogs and an epidemic-type aftershock sequence model}}.
\newblock \emph{\bibinfo{journal}{Phys. Rev. E}} \textbf{\bibinfo{volume}{99}},
  \bibinfo{pages}{042210} (\bibinfo{year}{2019}).

\bibitem{ogata2017statistics}
\bibinfo{author}{Ogata, Y.}
\newblock \bibinfo{title}{Statistics of earthquake activity: Models and methods
  for earthquake predictability studies}.
\newblock \emph{\bibinfo{journal}{Annu. Rev. Earth Planet. Sci.}}
  \textbf{\bibinfo{volume}{45}}, \bibinfo{pages}{497--527}
  (\bibinfo{year}{2017}).

\bibitem{papazachos1975foreshocks}
\bibinfo{author}{Papazachos, B.}
\newblock \bibinfo{title}{Foreshocks and earthquake prediction}.
\newblock \emph{\bibinfo{journal}{Tectonophysics}}
  \textbf{\bibinfo{volume}{28}}, \bibinfo{pages}{213--226}
  (\bibinfo{year}{1975}).

\bibitem{kagan1978statistical}
\bibinfo{author}{Kagan, Y.} \& \bibinfo{author}{Knopoff, L.}
\newblock \bibinfo{title}{Statistical study of the occurrence of shallow
  earthquakes}.
\newblock \emph{\bibinfo{journal}{Geophys. J. Int.}}
  \textbf{\bibinfo{volume}{55}}, \bibinfo{pages}{67--86}
  (\bibinfo{year}{1978}).

\bibitem{jones1979some}
\bibinfo{author}{Jones, L.~M.} \& \bibinfo{author}{Molnar, P.}
\newblock \bibinfo{title}{Some characteristics of foreshocks and their possible
  relationship to earthquake prediction and premonitory slip on faults}.
\newblock \emph{\bibinfo{journal}{J. Geophys. Res. Solid Earth.}}
  \textbf{\bibinfo{volume}{84}}, \bibinfo{pages}{3596--3608}
  (\bibinfo{year}{1979}).

\bibitem{console1993foreshock}
\bibinfo{author}{Console, R.}, \bibinfo{author}{Murru, M.} \&
  \bibinfo{author}{Alessandrini, B.}
\newblock \bibinfo{title}{Foreshock statistics and their possible relationship
  to earthquake prediction in the italian region}.
\newblock \emph{\bibinfo{journal}{Bull. Seismol. Soc. Am.}}
  \textbf{\bibinfo{volume}{83}}, \bibinfo{pages}{1248--1263}
  (\bibinfo{year}{1993}).

\bibitem{Peng-Vidale-Ishii-et-al-2007:seismicity}
\bibinfo{author}{Peng, Z.}, \bibinfo{author}{Vidale, J.~E.},
  \bibinfo{author}{Ishii, M.} \& \bibinfo{author}{Helmstetter, A.}
\newblock \bibinfo{title}{Seismicity rate immediately before and after main
  shock rupture from high-frequency waveforms in japan}.
\newblock \emph{\bibinfo{journal}{J. Geophys. Res. Solid Earth}}
  \textbf{\bibinfo{volume}{112}} (\bibinfo{year}{2007}).

\bibitem{gulia2019real}
\bibinfo{author}{Gulia, L.} \& \bibinfo{author}{Wiemer, S.}
\newblock \bibinfo{title}{Real-time discrimination of earthquake foreshocks and
  aftershocks}.
\newblock \emph{\bibinfo{journal}{Nature}} \textbf{\bibinfo{volume}{574}},
  \bibinfo{pages}{193--199} (\bibinfo{year}{2019}).

\bibitem{petrillo2021testing}
\bibinfo{author}{Petrillo, G.} \& \bibinfo{author}{Lippiello, E.}
\newblock \bibinfo{title}{Testing of the foreshock hypothesis within an
  epidemic like description of seismicity}.
\newblock \emph{\bibinfo{journal}{Geophys. J. Int.}}
  \textbf{\bibinfo{volume}{225}}, \bibinfo{pages}{1236--1257}
  (\bibinfo{year}{2021}).

\bibitem{reasenberg1992response}
\bibinfo{author}{Reasenberg, P.~A.} \& \bibinfo{author}{Simpson, R.~W.}
\newblock \bibinfo{title}{Response of regional seismicity to the static stress
  change produced by the loma prieta earthquake}.
\newblock \emph{\bibinfo{journal}{Science}} \textbf{\bibinfo{volume}{255}},
  \bibinfo{pages}{1687--1690} (\bibinfo{year}{1992}).

\bibitem{wyss2000change}
\bibinfo{author}{Wyss, M.} \& \bibinfo{author}{Wiemer, S.}
\newblock \bibinfo{title}{Change in the probability for earthquakes in southern
  california due to the landers magnitude 7.3 earthquake}.
\newblock \emph{\bibinfo{journal}{Science}} \textbf{\bibinfo{volume}{290}},
  \bibinfo{pages}{1334--1338} (\bibinfo{year}{2000}).

\bibitem{Lippiello2009a}
\bibinfo{author}{Lippiello, E.}, \bibinfo{author}{{De Arcangelis}, L.} \&
  \bibinfo{author}{Godano, C.}
\newblock \bibinfo{title}{{Role of static stress diffusion in the
  spatiotemporal organization of aftershocks}}.
\newblock \emph{\bibinfo{journal}{Phys. Rev. Lett.}}
  \textbf{\bibinfo{volume}{103}} (\bibinfo{year}{2009}).

\bibitem{tibi2003remote}
\bibinfo{author}{Tibi, R.}, \bibinfo{author}{Wiens, D.~A.} \&
  \bibinfo{author}{Inoue, H.}
\newblock \bibinfo{title}{Remote triggering of deep earthquakes in the 2002
  tonga sequences}.
\newblock \emph{\bibinfo{journal}{Nature}} \textbf{\bibinfo{volume}{424}},
  \bibinfo{pages}{921--925} (\bibinfo{year}{2003}).

\bibitem{Felzer}
\bibinfo{author}{Richards-Dinger, K.}, \bibinfo{author}{Stein, R.~S.} \&
  \bibinfo{author}{Toda, S.}
\newblock \bibinfo{title}{{Decay of aftershock density with distance does not
  indicate triggering by dynamic stress}}.
\newblock \emph{\bibinfo{journal}{Nature}} \textbf{\bibinfo{volume}{467}},
  \bibinfo{pages}{583--586} (\bibinfo{year}{2010}).

\bibitem{brodsky2014uses}
\bibinfo{author}{Brodsky, E.~E.} \& \bibinfo{author}{van~der Elst, N.~J.}
\newblock \bibinfo{title}{The uses of dynamic earthquake triggering}.
\newblock \emph{\bibinfo{journal}{Annu. Rev. Earth Planet. Sci.}}
  \textbf{\bibinfo{volume}{42}}, \bibinfo{pages}{317--339}
  (\bibinfo{year}{2014}).

\bibitem{pollitz201211}
\bibinfo{author}{Pollitz, F.~F.}, \bibinfo{author}{Stein, R.~S.},
  \bibinfo{author}{Sevilgen, V.} \& \bibinfo{author}{B{\"u}rgmann, R.}
\newblock \bibinfo{title}{The 11 april 2012 east indian ocean earthquake
  triggered large aftershocks worldwide}.
\newblock \emph{\bibinfo{journal}{Nature}} \textbf{\bibinfo{volume}{490}},
  \bibinfo{pages}{250--253} (\bibinfo{year}{2012}).

\bibitem{parsons2011absence}
\bibinfo{author}{Parsons, T.} \& \bibinfo{author}{Velasco, A.~A.}
\newblock \bibinfo{title}{Absence of remotely triggered large earthquakes
  beyond the mainshock region}.
\newblock \emph{\bibinfo{journal}{Nat. Geosci.}} \textbf{\bibinfo{volume}{4}},
  \bibinfo{pages}{312--316} (\bibinfo{year}{2011}).

\bibitem{parsons2014global}
\bibinfo{author}{Parsons, T.}, \bibinfo{author}{Segou, M.} \&
  \bibinfo{author}{Marzocchi, W.}
\newblock \bibinfo{title}{The global aftershock zone}.
\newblock \emph{\bibinfo{journal}{Tectonophysics}}
  \textbf{\bibinfo{volume}{618}}, \bibinfo{pages}{1--34}
  (\bibinfo{year}{2014}).

\bibitem{iwata2008low}
\bibinfo{author}{Iwata, T.}
\newblock \bibinfo{title}{Low detection capability of global earthquakes after
  the occurrence of large earthquakes: Investigation of the harvard cmt
  catalogue}.
\newblock \emph{\bibinfo{journal}{Geophys. J. Int.}}
  \textbf{\bibinfo{volume}{174}}, \bibinfo{pages}{849--856}
  (\bibinfo{year}{2008}).

\bibitem{garber2009predicting}
\bibinfo{author}{Garber, A.}, \bibinfo{author}{Hallerberg, S.} \&
  \bibinfo{author}{Kantz, H.}
\newblock \bibinfo{title}{Predicting extreme avalanches in self-organized
  critical sandpiles}.
\newblock \emph{\bibinfo{journal}{Phys. Rev. E}} \textbf{\bibinfo{volume}{80}},
  \bibinfo{pages}{026124} (\bibinfo{year}{2009}).

\bibitem{kossobokov2022}
\bibinfo{author}{Vladimir, G.~K.}, \bibinfo{author}{Anastasia, K.~N.} \& 
  \bibinfo{author}{Polina, D.~S}
\newblock \bibinfo{title}{{Seismic dynamics in advance of and after the largest earthquakes, 1985--2020}}.
\newblock \emph{\bibinfo{journal}{Surv. Geophys.}} \textbf{\bibinfo{volume}{43}},
  \bibinfo{pages}{423--436} (\bibinfo{year}{2022}).

\bibitem{Zhuang2010}
\bibinfo{author}{Zhuang, J.}, \bibinfo{author}{Werner, M.~J.},
  \bibinfo{author}{Harte, D.}, \bibinfo{author}{Hainzl, S.} \&
  \bibinfo{author}{Zhou, S.}
\newblock \bibinfo{title}{{Basic models of seismicity}}.
\newblock \emph{\bibinfo{journal}{Community Online Resour. Stat. Seism. Anal.}}
  \bibinfo{pages}{2--41} (\bibinfo{year}{2010}).

\bibitem{Zhuang2012}
\bibinfo{author}{Zhuang, J.}
\newblock \bibinfo{title}{{Long-term earthquake forecasts based on the
  epidemic-type aftershock sequence (ETAS) model for short-term clustering}}.
\newblock \emph{\bibinfo{journal}{Res. Geophys.}} \textbf{\bibinfo{volume}{2}},
  \bibinfo{pages}{8} (\bibinfo{year}{2012}).

\bibitem{Scholz2019}
\bibinfo{author}{Scholz, C.~H.}
\newblock \emph{\bibinfo{title}{The mechanics of earthquakes and faulting}}
  (\bibinfo{publisher}{Cambridge university press}, \bibinfo{year}{2019}).  
  
\bibitem{veen2008estimation}
\bibinfo{author}{Veen, A.} \& \bibinfo{author}{Schoenberg, F.~P.}
\newblock \bibinfo{title}{Estimation of space--time branching process models in
  seismology using an em--type algorithm}.
\newblock \emph{\bibinfo{journal}{J. Am. Stat. Assoc.}}
  \textbf{\bibinfo{volume}{103}}, \bibinfo{pages}{614--624}
  (\bibinfo{year}{2008}).

\bibitem{nandan2021seismicity}
\bibinfo{author}{Nandan, S.}, \bibinfo{author}{Ram, S.~K.},
  \bibinfo{author}{Ouillon, G.} \& \bibinfo{author}{Sornette, D.}
\newblock \bibinfo{title}{Is seismicity operating at a critical point?}
\newblock \emph{\bibinfo{journal}{Phys. Rev. Let.}}
  \textbf{\bibinfo{volume}{126}}, \bibinfo{pages}{128501}
  (\bibinfo{year}{2021}).

\end{thebibliography}


\end{document}